# Excitonic Resonant Emission−Absorption of Surface Plasmons in Transition Metal Dichalcogenides for Chip-Level Electronic−Photonic Integrated Circuits

Zhuan Zhu,[†,‡] Jiangtan Yuan,[§] Haiqing Zhou,[∥] Jonathan Hu,[⊥] Jing Zhang,[§] Chengli Wei,[⊥] Fang Yu,[∥] Shuo Chen,[∥] Yucheng Lan,[∥,#] Yao Yang,[§] Yanan Wang,[†,‡] Chao Niu,[⊥] Zhifeng Ren,[∥] Jun Lou,[§] Zhiming Wang,*[,†] and Jiming Bao*[,‡,†]

[†]Institute of Fundamental and Frontier Sciences, University of Electronic Science and Technology of China, Chengdu, Sichuan 610054, China

[‡]Department of Electrical and Computer Engineering and [∥]Department of Physics and Texas Center for Superconductivity, University of Houston, Houston, Texas 77204, United States

[§]Department of Materials Science and NanoEngineering, Rice University, Houston, Texas 77005, United States

[⊥]Department of Electrical & Computer Engineering, Baylor University, Waco, Texas 76798, United States

[#]Department of Physics and Engineering Physics, Morgan State University, Baltimore, Maryland 21251, United States

**ABSTRACT:** The monolithic integration of electronics and photonics has attracted enormous attention due to its potential applications. A major challenge to this integration is the identification of suitable materials that can emit and absorb light at the same wavelength. In this paper we utilize unique excitonic transitions in $WS_2$ monolayers and show that $WS_2$ exhibits a perfect overlap between its absorption and photoluminescence spectra. By coupling $WS_2$ to Ag nanowires, we then show that $WS_2$ monolayers are able to excite and absorb surface plasmons of Ag nanowires at the same wavelength of exciton photoluminescence. This resonant absorption by $WS_2$ is distinguished from that of the ohmic propagation loss of silver nanowires, resulting in a short propagation length of surface plasmons. Our demonstration of resonant optical generation and detection of surface plasmons enables nanoscale optical communication and paves the way for on-chip electronic−photonic integrated circuits.

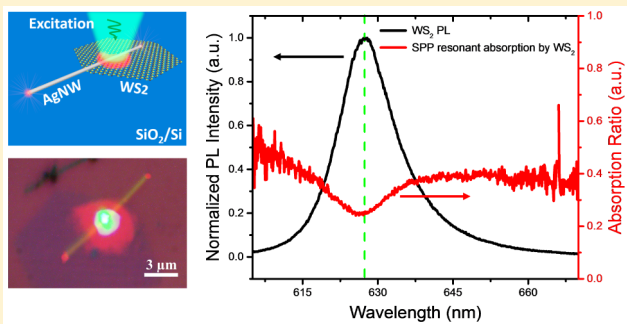

The emission and absorption of photons through electronic transitions is a basic light−matter interaction and forms the foundation for optoelectronic devices and optical tele-communications. Because optical emission and absorption originate from the same dipole transition matrix element in quantum mechanics, they are reversible and have the same magnitude in principle. In other words, if a photon is emitted by a quantum system, it should also be absorbed by the same system. Such emission−absorption resonance is a characteristic of two-level model systems and is ubiquitous in atomic spectra. However, such a resonance is hard to observe in conventional optical materials such as III−V semiconductors. In such a material, a photon is emitted at the band gap, but the absorption by the same material is virtually zero because the absorption increases as the square root of $E − E_g$, where $E_g$ is the band gap. Such a resonance is also difficult to observe in semiconductor quantum dots, artificial atoms that exhibit atom-like discrete energy levels and absorption spectra: there is typically a significant red-shift of the emission peak relative to the absorption spectrum.[1] This Stokes shift is a general optical property of quantum dots; it cannot be eliminated simply by changing the size of quantum dots.[2−4] The resonance, however, can be achieved under certain conditions by carefully tuning the Fermi level of QDs with an electrical gating.[5]

The ability of an electronic material to emit and absorb photons at the same wavelength is essential to chip-scale optical communication and monolithic integration of electronic and photonic circuits, which would not only increase the speed of computation with reduced power consumption but also provide new applications such as quantum information processing and chemical or biological sensing.[6−9] The idea of an electronic−photonic integrated circuit (EPIC) dates back decades after the great successes of silicon microelectronics and fiber-based

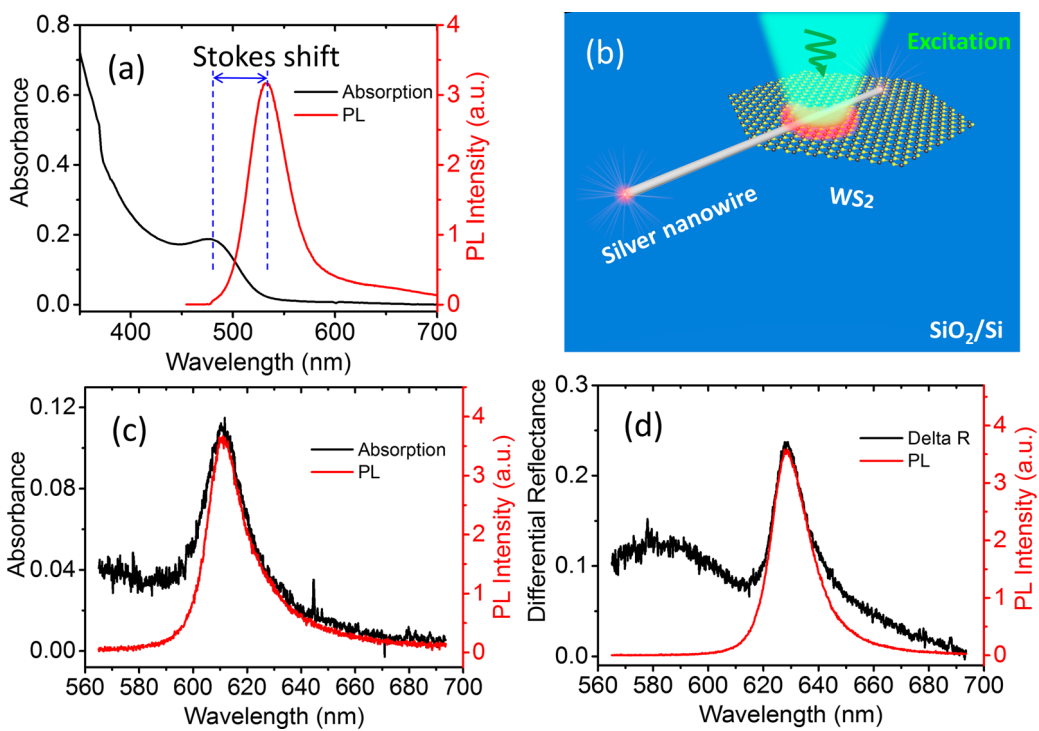

**Figure 1.** Photoluminescence, absorption, and experimental setup. (a) Photoluminescence and absorption spectra of CdTe quantum dots. (b) Experimental configuration showing the photoexcitation of $WS_2$ and the coupling between the Ag nanowire and $WS_2$. (c and d) Photoluminescence and absorption spectra of $WS_2$ monolayers on (c) a quartz and (d) a $SiO_2/Si$ substrate. Differential reflectance is equivalent to the absorption spectrum for atomically thin films.

optical communications. Silicon is a natural choice of material for EPIC because of silicon-based mature electronics, but silicon is an indirect band gap semiconductor, which makes it challenging to use Si for efficient light generation and detection. In silicon photonics, much effort has been made to either transform silicon to an active optical material or integrate silicon with other optically active materials. Decades of research has culminated in the recent demonstration of a board-level EPIC.[10] However, that technique cannot be simply used to build a monolithic chip-level EPIC due to potential material incompatibility and scale mismatch between electronics and photonics. This nanometer versus micrometer scale mismatch problem has now been solved by the emergence of plasmonics, which enables confinement of light to a size compatible with electronics.[11−14] Much progress has since been made toward the chip-level EPIC. Falk et al. demonstrated local electrical detection of surface plasmon polaritons (SPPs) using Ge nanowires;[15] Neutens et al. utilized an integrated GaAs photodetector to detect gap SPPs.[16] Direct SPP generation by electricity has also been realized through nanometer-scale light-emitting diodes using III−V semiconductors.[17−19]

Despite enormous efforts and significant progress, chip-level optical communication, i.e., the simultaneous generation and detection of SPPs, still remains a challenge. In fact, these conventional semiconductors are not the right material platform to fabricate monolithic chip-level EPIC. Taking the above-mentioned SPP devices as an example, although an SPP can be directly generated by a nanoscale integrated GaAs diode,[17,18] it can hardly be detected using the same GaAs because optical absorption at the band edge of GaAs is negligible. Two-dimensional transition metal dichalcogenides (TMDs) have recently emerged as promising nanomaterials for both electronic and photonic device applications.[20−26] TMDs such as $MoS_2$ have been used to generate or detect SPPs, but other than being atomically thin, their unique advantage of being able to emit and absorb photons at the same wavelength has not been demonstrated compared to the above-mentioned semiconductors.[27−29] Moreover, multiple exciton emissions from $MoS_2$ make the coupling of $MoS_2$ photoluminescence to SPPs of Ag waveguides unnecessarily complicated, which could hinder future device applications.

An example of Stokes shift is shown in Figure 1a. The difference in peak positions between absorption and photoluminescence spectra of ∼2 nm CdTe quantum dots is significant such that the absorption at the photoluminescence peak drops to near zero, although both absorption and photoluminescence spectra exhibit the characteristic of exciton transition. This is very different for monolayer $WS_2$.[26] Figure 1b shows the experimental setup. Large-size monolayers of $WS_2$ are synthesized by chemical vapor deposition on $SiO_2/Si$ or quartz substrates.[30] Ag nanowires (AgNWs) with a diameter of ∼120 nm (AgNWs-120 from ACSMATERIAL) are used as plasmonic waveguides. A 532 nm laser is used to photoexcite the monolayer $WS_2$ through a microscope objective. For $WS_2$ on the $SiO_2/Si$ substrates, the differential reflectance is measured to obtain its absorption spectrum.[31,32] Figure 1c,d show nearly perfect spectral overlaps between photoluminescence and absorption spectra of $WS_2$ monolayers on either a quartz or $SiO_2/Si$ substrate. Close peak positions between absorbance and photoluminescence of $WS_2$ monolayers were also reported by Zhao[32] and Peimyoo,[33] but a bigger Stokes shift was observed by Eda and Maier.[26] Similar discrepancies among different groups can also be found for $MoS_2$,[26,31] and they are believed to be due to different crystal quality and doping levels of monolayer TMDs. Note that the shorter peak wavelength in

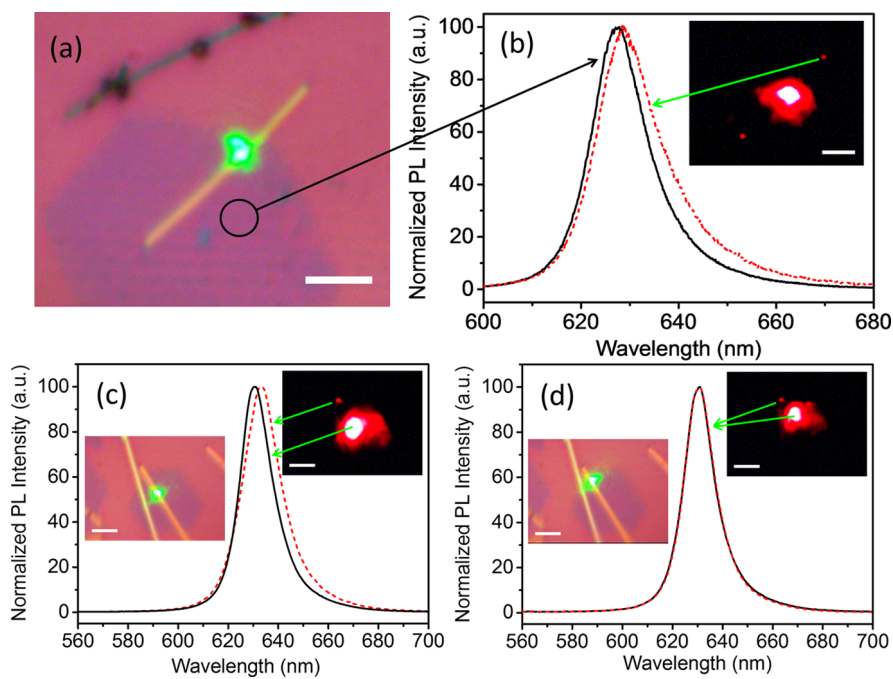

**Figure 2.** Spectrum of surface plasmon polaritons (SPPs) created by the photoluminescence (PL) of $WS_2$. (a) Photoexcitation image of one AgNW/$WS_2$ device. (b) Photoluminescence (black) of $WS_2$ away from the Ag nanowire and spectrum of the SPPs scattered from the silver nanowire tip (red). Inset is the associated luminescence image of the excitation image in (a). (c and d) Spectra of emitted light from the photoexcitation spot (black) and from the silver nanowire tip (red) when another device is photoexcited on AgNW/$WS_2$ (c) nearly in the middle or (d) at the edge of the $WS_2$ crystal. Insets are excitation images and the associated luminescence images. Scale bars: 3 $\mu m$. Laser wavelength: 532 nm.

Figure 1c than in Figure 1d was also observed by Peimyoo et al. due to substrate-induced strain.[33,34]

Having determined the peak wavelength of photoluminescence and established the resonance relationship between emission and absorption, we are in a position to investigate how the photoluminescence of $WS_2$ is coupled to surface plasmon polaritons and how the SPPs are subsequently absorbed by $WS_2$. Using the configuration in Figure 1b, the generation and propagation of SPPs are confirmed in Figure 2 by the appearance of red light from the ends of the Ag nanowires. Note that unlike the $MoS_2$ work, the photoluminescence spectrum remains the same no matter whether the photoexcitation is performed on bare $WS_2$ away from the nanowire or on the AgNW/$WS_2$.[27,28,35] Hence, we do not distinguish between the photoluminescence spectra from a bare $WS_2$ and from AgNW/$WS_2$, and we use either of them for convenience. On the other hand, it is not accurate to simply take the photoluminescence spectrum from AgNW/$WS_2$ as the SPP spectrum of Ag nanowires because a significant amount of photoluminescence originates directly from $WS_2$. In other words, the spectrum of coupled SPPs should be measured from the end emission of Ag nanowires, which was also used to determine the spectrum of SPPs coupled to $MoS_2$ monolayers. However, a clear red shift of SPPs can be seen in Figure 2b,c compared to the spectrum of $WS_2$, in contrast to previous work in $MoS_2$.[27,28,35]

There is always a spectral shift between the emission from the AgNW/$WS_2$ photoexcitation spot and the SPPs scattered at the AgNW end as long as the AgNW/$WS_2$ is excited inside the $WS_2$ monolayer. It is not until the photoexcitation is moved to the farthest edge of $WS_2$ and the SPPs' travel distance is short enough that the two spectra become almost identical, as shown in Figure 2d. On the basis of these observations, we can now conclude that the SPP has the same spectrum as $WS_2$ photoluminescence when it is initially created by the photoexcitation of AgNW/$WS_2$. The observations from Figure 2b,c indicate that the spectral red-shift can be induced by either a long AgNW waveguide or a combination of AgNW and $WS_2$. To find out their relative contributions, we use the same device shown in Figure 2a and move the photoexcitation spot more into $WS_2$. Figure 3a,b show the SPP travels toward both ends 1A and 1B of the nanowire. Despite a longer travel distance to 1B, the SPP from 1A experiences more red-shift than that from 1B, as revealed in Figure 3e. This observation leads us to conclude that SPP interaction with $WS_2$ has a more significant contribution to its spectral red-shift.

The effect of $WS_2$ and Ag nanowire on the spectral red-shift of SPPs can also be seen in Figure 3b, c, d, and f, when $WS_2$ is photoexcited at different locations along the Ag nanowire. A larger spectral shift is observed when the propagation distance along AgNW/$WS_2$ increases. Moreover, the spectral shift is accompanied by a decreasing intensity of the SPPs. Because the photoexcitation must be performed over AgNW/$WS_2$, simultaneous propagation losses from both the Ag nanowire and $WS_2$ cannot be avoided. In order to distinguish these two propagation losses and investigate their underlying loss mechanisms, we normalize the spectrum of SPPs with respect to the photoluminescence spectrum of $WS_2$. As can be seen from Figure 3g, both SPP spectra exhibit a gradual decrease in spectral loss at longer wavelengths and a resonant absorption, but the resonant absorption in spectrum 1A is more pronounced than that of 1B. We attribute the broad absorption to the SPP propagation loss from AgNW/$WS_2$[36] and the resonant absorption to $WS_2$. This conclusion agrees with the absorption spectrum of $WS_2$ in Figure 1, where a strong resonant absorption peak is superposed on the broadband background. The weaker $WS_2$ absorption for 1B SPP is due to its shorter interaction length with $WS_2$. To obtain a better resonant absorption spectrum with less broadband absorption background, we normalized the spectrum of 3B using the spectrum of 1B as a reference; the result in Figure 3h shows a clear resonance with the reduced propagation loss from the nanowire. We also include the photoluminescence in the same plot, and an excellent overlap is observed. Thus, we can conclude that the SPP is resonantly absorbed by WS$_2$ at the same wavelength as the photoluminescence; there is no difference between the SPP and free-space light when it interacts with WS$_2$; over a short distance of less than 2 $\mu$m, the propagation loss of the SPP is dominated by resonant absorption of WS$_2$.

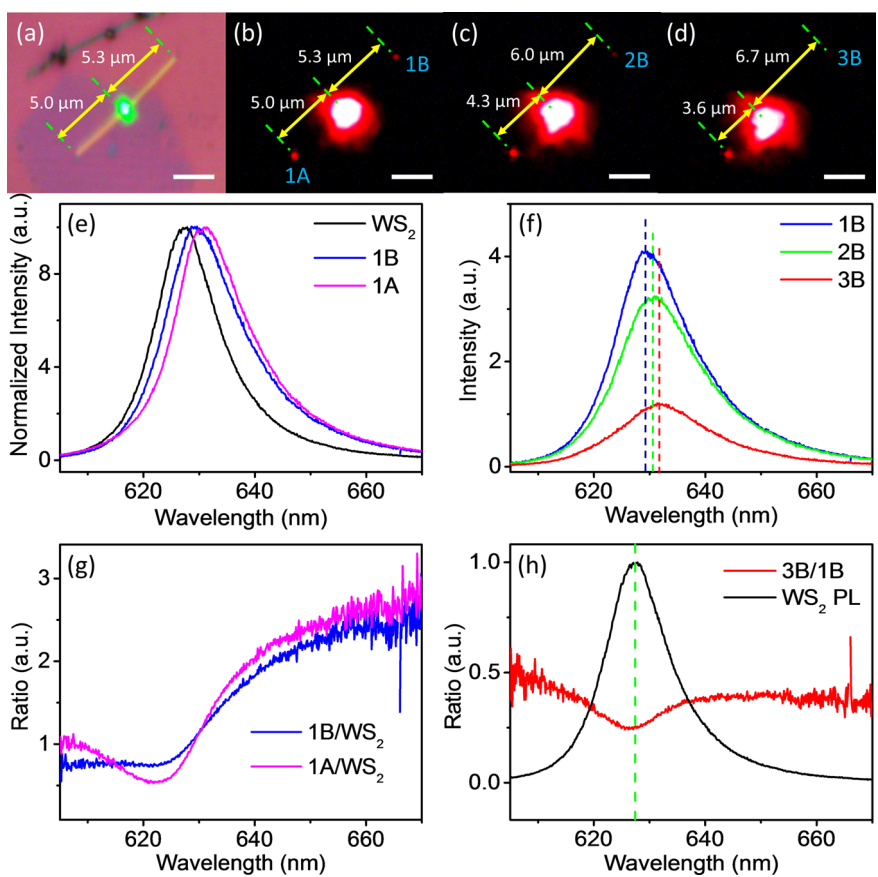

**Figure 3.** Absorption spectrum of surface plasmon polaritons (SPPs) by WS$_2$. (a) Excitation image of the device. (b) Luminescence image of (a). (c and d) Luminescence images when WS$_2$ is photoexcited at two other locations along the Ag nanowire. (e) Normalized spectra of the photoexcitation (black) and SPPs from points 1A and 1B in part (b). (f) SPP spectra from 1B, 2B, and 3B in parts (b)−(d). (g) Ratios of the spectra 1B and 1A to the exciton spectrum of WS$_2$. (h) Ratio of spectrum 3B to spectrum 1B (red), in comparison to the spectrum of WS$_2$. Scale bars: 3 $\mu$m.

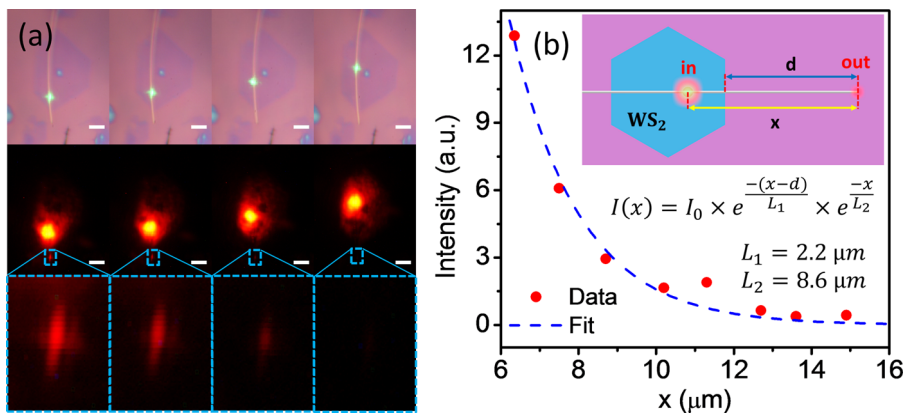

**Figure 4.** Measurement of SPP propagation length. (a) Photoexcitation (top) and SPP emission images (zoomed-in images at the bottom). (b) Intensity of the SPP based on (a) as a function of propagation length. Inset: Schematic of the experimental design. Scale bars: 5 $\mu$m.

To quantify the resonant absorption of SPP by WS$_2$, we selected a device with a longer Ag nanowire both on and outside of WS$_2$ so that the SPP propagation distance can be varied in a wider range. Figure 4 shows the experimental design and calculation principle. When an SPP is created by photo-excitation, it will go through two regions before reaching the clear end of the nanowire waveguide: the first region is the nanowire on WS$_2$, and the second one is the nanowire alone. We designate $\alpha_1$ and $\alpha_2$ as the absorption coefficients due to WS$_2$ and the nanowire, respectively. The total absorption coefficient when the nanowire on WS$_2$ can be approximated as $\alpha_1 + \alpha_2$. Thus, the intensity of SPP end emission at the point $x$ can be expressed as

$$I(x) = I_0 e^{-(\alpha_1+\alpha_2)(x-d)} e^{-\alpha_2 d} = I_0 e^{-\alpha_1(x-d)} e^{-\alpha_2 x} \quad (1)$$

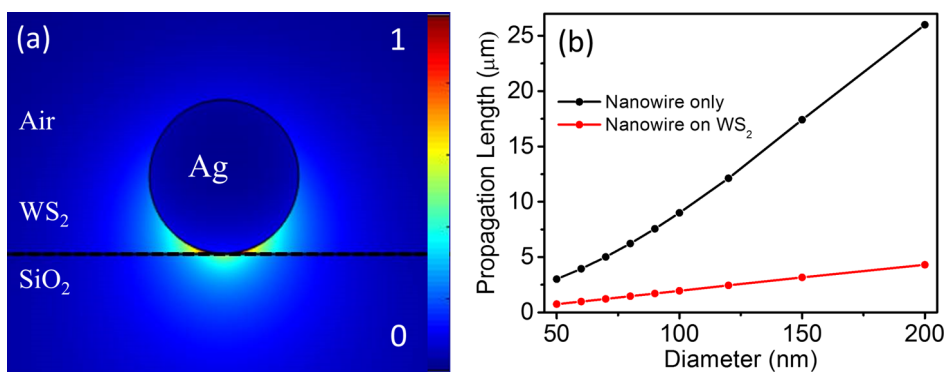

**Figure 5.** Numerical calculation of SPP propagation length. (a) Cross-sectional view of a field intensity profile of a propagating SPP at 633 nm. The diameter of the nanowire is 120 nm. (b) Calculated SPP propagation lengths along the Ag nanowire as a function of the nanowire diameter with and without WS$_2$.

Here $I_0$ is the intensity of the exciton SPP at the point of photoexcitation. If we define propagation length $L_1$ and $L_2$ as $L_1 = 1/\alpha_1$, $L_2 = 1/\alpha_2$, then the equation can be written as

$$I(x) = I_0 e^{-(\alpha_1+\alpha_2)(x-d)} e^{-\alpha_2 d} = I_0 e^{-(x-d)/L_1} e^{-x/L_2} \quad (2)$$

The propagation lengths $L_1$ and $L_2$ can be obtained by curve fitting eq 2. Figure 4a shows a series of snapshots of SPP photoexcitation and SPP emission. The integrated SPP intensity is plotted in Figure 4b as a function of SPP travel distance. Curve fitting by eq 2 gives us $L_1 = 2.2 \ \mu$m and $L_2 = 8.6 \ \mu$m. The propagation length $L$ of SPP with the nanowire on WS$_2$ is ~1.8 $\mu$m based on the total absorption of $\alpha_1 + \alpha_2$. The propagation $L_2$ is similar to previously reported values using silver nanowires with a similar diameter.[36−40] We also measured the propagation length of a silver nanowire fully placed on a WS$_2$ crystal. In that case, we obtain $L_1 = 2.4 \ \mu$m, $L_2 = 12.2 \ \mu$m, and the propagation length $L = 2.0 \ \mu$m.

The understanding of SPP absorption allows us to numerically calculate the propagation length based on the measured optical constants of WS$_2$.[41] Note that the strong absorption of SPPs by a WS$_2$ monolayer is also a result of plasmon-enhanced strong light−matter interaction. Figure 5a shows the local field distribution around the AgNW/WS$_2$ at 633 nm, which is close to the wavelength of WS$_2$ SPPs. This is the fundamental mode of an SPP, and it has been previously studied without a WS$_2$ monolayer.[39,40,42] Although only one monolayer thick, the effect of monolayer WS$_2$ is noticeable: the field becomes more concentrated in the region where the nanowire and WS$_2$ overlap, leading to a short propagation length. Figure 5b shows the dependence of SPP propagation length on nanowire diameter with and without WS$_2$. It can be seen that the effect of WS$_2$ is strong regardless of the nanowire's diameter. For the 120 nm nanowire, we obtain a propagation length of 2.4 $\mu$m, which is in good agreement with the experimental value.

The well-defined photoluminescence and absorption peaks as well as their excellent spectral overlap are a result of excitonic transitions in WS$_2$ monolayers and are a general property of monolayer TMDs.[26,31,33,43,44] A large exciton binding energy in monolayer TMDs enables a discrete absorption spectrum.[31,43,44] The SPP wavelength can be selected by choosing different TMDs or tuned by straining or doping.[34] Here chemically synthesized Ag nanowires are used only to demonstrate the concept, and lithographically defined plasmonic waveguides and couplers could be used for large-scale integration.[19,38]

In conclusion, we have demonstrated optical emission and absorption resonance not only for free-space light but also for surface plasmon polaritons. The same resonant wavelength of photon emission and absorption is determined by direct exciton transitions in monolayer WS$_2$. Although the generation and detection of SPPs were demonstrated optically, the device can be readily modified to perform local generation and detection of SPPs by electrical means, i.e., nanoscale optical communication between transistors on the same chip. Because the planar structure of 2D TMDs is compatible with the current lithographic fabrication process, we expect the demonstration of a complete EPIC in the not too distant future.

## ■ AUTHOR INFORMATION

**Corresponding Authors**
*E-mail: jbao@uh.edu.
*E-mail: zhmwang@uestc.edu.cn.

**Notes**
The authors declare no competing financial interest.

## ■ ACKNOWLEDGMENTS

J.M.B. acknowledges support from the National Science Foundation (CAREER Award ECCS-1240500) and the Robert A. Welch Foundation (E-1728), J.L. acknowledges support from the Welch Foundation grant C-1716 and NSF ECCS-1327093, Z.F.R. acknowledges support from the Department of Energy DE-FG02-13ER46917/DE-SC0010831 and US Defense Threatening Reduction Agency (DTRA) under grant FA 7000-13-1-0001. J.H. acknowledges support from the Vice Provost for Research at Baylor University.

F